\documentclass[aps,prl,showpacs,amsmath,amssymb,twocolumn,10pt]{revtex4-1} 
\usepackage{amsthm}
\usepackage{dsfont}
\usepackage[dvips]{graphicx}

\newcommand\ZZ{{\mathds{Z}}}
\newcommand\RR{{\mathds{R}}}

\begin{document} 

\title{Phase transitions detached from stationary points of the energy landscape}  

\author{Michael Kastner} 
\email{kastner@sun.ac.za} 
\affiliation{National Institute for Theoretical Physics (NITheP), Stellenbosch 7600, South Africa} 
\affiliation{Institute of Theoretical Physics,  University of Stellenbosch, Stellenbosch 7600, South Africa}

\author{Dhagash Mehta} 
\email{dbmehta@syr.edu} 
\affiliation{Department of Physics, Syracuse University, Syracuse, New York 13244, USA} 

\date{\today}
 
\begin{abstract}
The stationary points of the potential energy function $V$ are studied for the $\phi^4$ model on a two-dimensional square lattice with nearest-neighbor interactions. 
On the basis of analytical and numerical results, we explore 
the relation of stationary points to the occurrence of thermodynamic phase transitions. We find that the phase transition potential energy of the $\phi^4$ model does in general not coincide with the potential energy of any of the stationary points of $V$. This disproves earlier, allegedly rigorous, claims in the literature on necessary conditions for the existence of phase transitions. Moreover, we find evidence that the indices of stationary points scale extensively with the system size, and therefore the index density can be used to characterize features of the energy landscape in the infinite-system limit. We conclude that the finite-system stationary points provide one possible mechanism of how a phase transition can arise, but not the only one.
\end{abstract}


\maketitle 

The stationary points of the potential energy function or other classical energy functions can be employed to calculate or estimate physical quantities. Well-known examples include transition state theory or Kramers's reaction rate theory for the thermally activated escape from metastable states, where the barrier height (corresponding to the difference between potential energies at certain stationary points of the potential energy function) plays an essential role. More recently, a large variety of related techniques has become popular under the name of energy landscape methods \cite{Wales}, with applications to many-body systems as diverse as metallic clusters, or biomolecules and their folding transitions. While the mentioned applications focus mostly on the numerical investigation of finite systems, the analysis of stationary points has also proved useful for analytical studies of $N$-body systems in the thermodynamic limit. One field of research where such methods have been fruitfully applied is disordered systems undergoing a dynamical glass transition \cite{KurchanLaloux96,*CavagnaGiardinaParisi98,*CavagnaGarrahanGiardina99}.

Another line of research based on stationary points but focusing on equilibrium phase transitions in the thermodynamic limit $N\to\infty$, dates back to about the same time \cite{CaCaClePe97,*CaPeCo00}. This approach, originally formulated in terms of topology changes of configuration space submanifolds, can be rephrased in terms of stationary points of the potential energy function $V$, i.e.\ configuration space points $q^\text{s}$ satisfying $\nabla V(q^\text{s})=0$. The underlying idea can be understood as follows \cite{CasettiNardiniNerattini11}: Thermodynamic equilibrium properties are encoded in the thermodynamic limit value of the microcanonical configurational entropy
\begin{equation}\label{eq:se}
s_N(v)=\frac{1}{N}\ln\int_\Gamma \delta[V(x)-Nv]\mathrm{d}x = \frac{1}{N}\ln\int_{\Sigma_v}\frac{\mathrm{d}\Sigma}{|\nabla V|},
\end{equation}
where $\Gamma$ denotes configuration space and $\mathrm{d}x$ its volume measure, $\Sigma_v\subset\Gamma$ is the hypersurface of constant-potential energy $V=Nv$, and $\mathrm{d}\Sigma$ stands for the $(N-1)$-dimensional Hausdorff measure on $\Sigma_v$. At a stationary point, we have $\nabla V=0$, the integrand on the righthand side of \eqref{eq:se} diverges, and we may expect the stationary point to give an important contribution to the integral. Indeed, it has been shown that, for finite $N$, every stationary point $q^\text{s}$ of $V$ induces nonanalytic behavior in $s_N(v)$ precisely at the potential energy of the stationary point, $v=V(q^\text{s})/N$ \cite{KaSchneSchrei07,*KaSchneSchrei08}.

Nonanalyticities of thermodynamic functions are hallmarks of phase transitions. Having observed that stationary points of $V$ cause nonanalyticities in the finite-system entropy $s_N$, it seems natural to ask whether they may also be responsible for nonanalytic behavior in the infinite-system entropy, i.e., for the occurrence of phase transitions. While for finite $N$ every stationary point of $V$ induces a nonanalyticity in $s_N$, the majority of the nonanalyticities does not survive the thermodynamic limit. Two questions turn out to be central to the understanding of these observations. (1) Under what conditions can the nonanalytic behavior of $s_N$, induced by a stationary point of $V$, survive the thermodynamic limit? A possible scenario has been depicted in \cite{KaSchne08,*KaSchneSchrei08}, and it turns out that the Hessian determinant of the potential energy function, evaluated at the stationary points, is crucial for discriminating whether or not the stationary points can induce a phase transition in the thermodynamic limit. For some models this insight has proved particularly useful in that it facilitates the analytic computation of phase transition energies even in the absence of an exact thermodynamic solution \cite{NardiniCasetti09,*Kastner11}. (2) Are stationary points necessary for a phase transition to take place? Or is there another mechanism, distinct from the one sketched above, which can give rise to a phase transition? This question was addressed by a theorem in \cite{FraPe04,*FraPeSpi07}, claiming that, for a large class of systems with short-range interactions, the presence of stationary point is indeed necessary for a phase transition to occur.

The model study presented in this Letter puts this alleged relation of stationary points of the potential energy function $V$ and thermodynamic phase transitions to the test. We apply three different methods to extract information about the stationary points of the potential energy function of the two-dimensional nearest-neighbor $\phi^4$ model. This model is known to have a continuous phase transition which is in the universality class of the two-dimensional Ising model but, in contrast to that model, is amenable to energy landscape methods by virtue of its continuous configuration space. The results from two complementary numerical techniques, as well as a rigorous analytical upper bound, all provide evidence that stationary points occur only at nonpositive potential energies.

The implications of this finding are significant. Among other things, they show by counterexample that the theorem on the relation between stationary points and thermodynamic phase transitions, allegedly proven in \cite{FraPe04,*FraPeSpi07}, is incorrect. Instead, we observe that, even for short-range interacting systems, thermodynamic phase transitions can occur at energies not related to any stationary points. As a consequence, further mechanisms of how phase transitions arise, possibly not related to finite-system stationary points, must exist, and we will comment on possible scenarios towards the end of this Letter.

{\em Two-dimensional nearest-neighbor $\phi^4$ model.---}On a finite square lattice $\Lambda\subset\ZZ^2$ consisting of $N=L^2$ sites, a real degree of freedom $\phi_i$ is assigned to each lattice site $i\in\Lambda$. By $\mathcal{N}(i)$ we denote the subset of $\Lambda$ consisting of the four nearest-neighboring sites of $i$ on the lattice under the assumption of periodic boundary conditions. The potential energy function of this model is given by
\begin{equation}\label{eq:V}
V(q)=\sum_{i\in\Lambda}\Biggl[\frac{\lambda}{4!}q_i^4-\frac{\mu^2}{2}q_i^2+\frac{J}{4}\sum_{j\in\mathcal{N}(i)}(q_i-q_j)^2\Biggr],
\end{equation}
where $q=(q_1,\dots,q_N)$ denotes a point in configuration space $\Gamma=\RR^N$
\footnote{Our definition of $V$ coincides with the one in \cite{FraCaSpiPe99}, but differs from \cite{FraPeSpi00} by a factor $1/6$ in the quartic term. Judging from the critical temperatures and energies reported in the latter, as well as from their reference to \cite{FraCaSpiPe99}, we assume that there is a misprint in \cite{FraPeSpi00}. Note that for the main conclusions of the present Letter the precise values of any of the constants do not matter.}%
.
$J>0$ determines the coupling strength between nearest-neighboring sites and the parameters $\lambda,\mu>0$ characterize a local double-well potential experienced by each degree of freedom.

In the thermodynamic limit $N\to\infty$ this model is known to undergo, at some critical temperature $T_\text{c}$, a continuous phase transition, in the sense that the configurational canonical free energy $f(T)$
is nonanalytic at $T=T_\text{c}$. The transition is from a ``ferromagnetic'' phase with nonzero average particle displacement to a ``paramagnetic'' phase with vanishing average displacement (see \cite{MilchevHeermannBinder86} for more details as well as for Monte Carlo results). 

\begin{figure}\center
\includegraphics[width=0.5\linewidth]{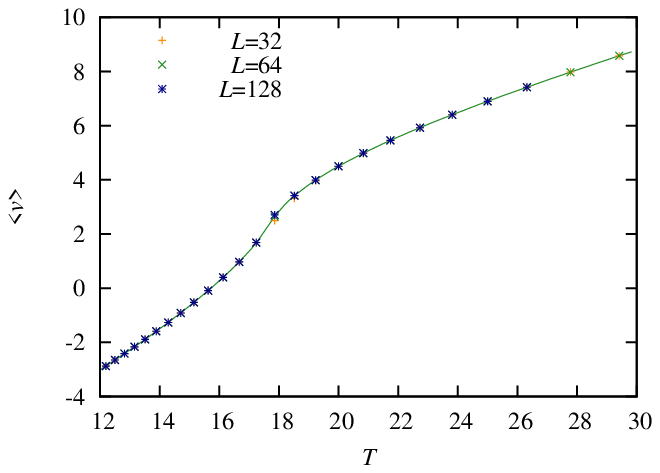}
\hspace{-0.03\linewidth}
\includegraphics[width=0.5\linewidth]{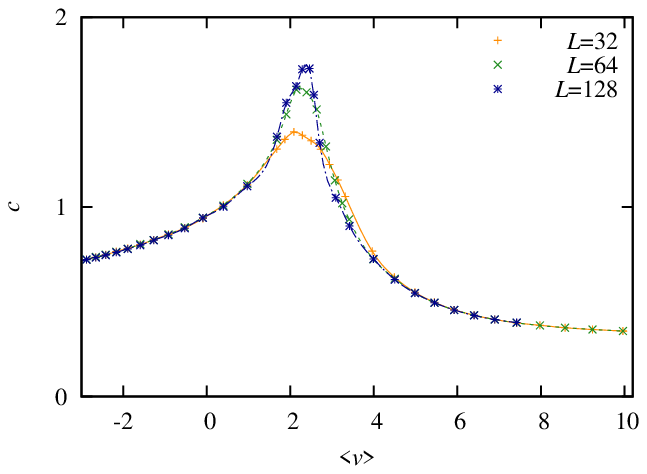}
\caption{\label{fig:MC}
(Color online) Monte Carlo results for the two-dimensional $\phi^4$ model \eqref{eq:V} with $\lambda=3/5$, $\mu^2=2$, and $J=1$. System sizes $N=L\times L$ are plotted with $L$ ranging from 32 to 128. Left panel: the canonical average $\langle v\rangle$ of the potential energy per lattice site as a function of the temperature $T$. Right panel: The canonical specific heat $c=(\langle V^2\rangle-\langle V\rangle^2)/(NT^2)$, plotted as a function of $\langle v\rangle$. The lines between the data points serve as a guide to the eye.
}
\end{figure}

Instead of $T_\text{c}$, it is more adequate for our purposes to compare to the critical potential energy per lattice site $v_\text{c}$ of the transition \cite{Kastner06}. Both quantities are unambiguously related to each other in the thermodynamic limit via the caloric curve $v(T)$, independently of the statistical ensemble used. The critical potential energy $v_\text{c}$ is less frequently studied, in fact the only data we could find in the literature are from Monte Carlo simulations of fairly small system sizes $N=20\times20$ in \cite{FraCaSpiPe99}, with parameter values $\lambda=3/5$, $\mu^2=2$, and $J=1$. We mostly use the same parameter values in the following. Since the value of $v_\text{c}$ is a crucial benchmark when relating our stationary point analysis to the phase transition of the $\phi^4$ model, we have performed standard Metropolis Monte Carlo simulations for somewhat larger system sizes up to $128\times128$ and $10^7$ lattice sweeps. We have sampled, among other observables, the canonical average $\langle v\rangle$ of the potential energy per particle, and plots of some of the Monte Carlo results are shown in Fig.\ \ref{fig:MC}. The quality of the data is sufficient to establish, beyond any reasonable doubt, a critical potential energy $v_\text{c}\approx2.2$ well above zero.

{\em Numerical continuation method.---}For $J=0$, i.e., in the absence of coupling, the stationary points $q^\text{s}$ of the potential $V$ in \eqref{eq:V} can be calculated analytically without difficulty, obtaining the $3^N$ solutions $q^\text{s}=(q^\text{s}_1,\dotsc,q^\text{s}_N)$ with $q^\text{s}_j\in\{0,\pm\sqrt{6\mu^2/\lambda}\}$. Knowledge of these solutions permits us to continue them to $J>0$ by numerical continuation; see \cite{Mehta_etal09,*Mehta11} for a description of the homotopy continuation method we have actually been using. Under certain conditions on the initial (decoupled) and final (coupled) potentials, this method is known to yield all stationary points of $V$. However, since the number of stationary points for $J=0$ grows exponentially with the number of lattice sites, such an analysis is restricted to fairly small system sizes.

\begin{figure}\center
\includegraphics[width=0.5\linewidth]{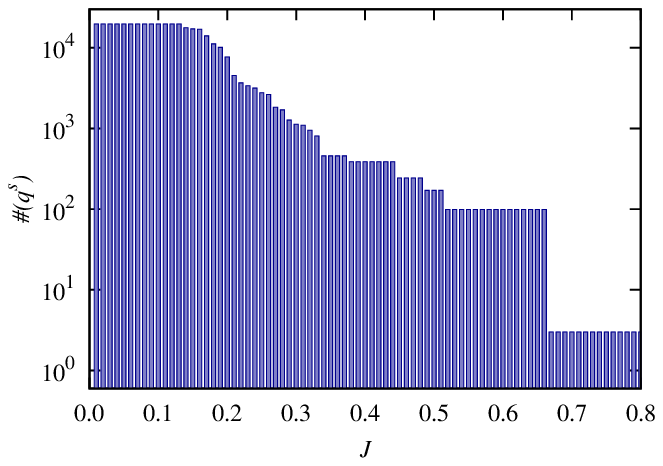}
\hspace{-0.03\linewidth}
\includegraphics[width=0.5\linewidth]{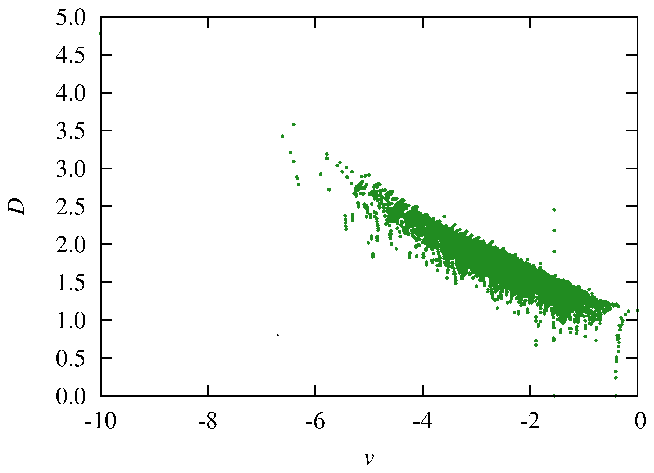}
\caption{\label{fig:cont}
(Color online) Numerical results from the homotopy continuation method. Left panel: The number of stationary points of $V$ for $N=3\times3$, plotted logarithmically as a function of the coupling $J$. Right panel: For all 10\,288\,973 stationary points $q^\text{s}$ of a $4\times4$ lattice with $J=0.2$, the scaled Hessian determinant $D=|\det{\mathcal H}_V(q^\text{s})|^{1/N}$ is shown vs the stationary value $v^\text{s}=V(q^\text{s})/N$, demonstrating that $v^\text{s}\leq0$ for all $q^\text{s}$.
}
\end{figure}

We have used the homotopy continuation method to compute all stationary points of $V$ for various values of $J$ and system sizes up to $4\times4$, and the results show the following features. First, upon increasing the coupling constant $J$, the number $\#(q^\text{s})$ of stationary points dramatically decreases from $3^N$ for $J=0$ to only 3 stationary points for larger $J$. This behavior is illustrated for $N=3\times3$ in the left panel of Fig.\ \ref{fig:cont}. Second, the stationary values $v^\text{s}=V(q^\text{s})/N$, i.e., the potential energy per lattice site evaluated at a stationary point, is found to be never larger than zero. This is illustrated for $N=4\times4$ in the right panel of Fig.\ \ref{fig:cont}. 

{\em Newton-Raphson method.---}We use Monte Carlo dynamics in configuration space to generate a large set of initial states, and apply the Newton-Raphson method to these initial states to find stationary point of $V$. In contrast to the homotopy continuation method, there is in general no way of knowing whether all stationary points have been found. An advantage, however, is that the Newton-Raphson method can be applied to system sizes much larger than $4\times4$.

We have used the routine {\tt newt} from \cite{NumRecC}, a globally convergent version of the Newton-Raphson method, to compute stationary points of $\phi^4$ lattices of sizes up to $32\times32$. For small $J$, the number of stationary points for such system sizes is huge and only a tiny fraction of them can be tracked down. For sufficiently large $J$ \footnote{Numerically, we find that, for a given $N$, there is a $J_0(N)$ such that only three stationary points exist for all $J>J_0(N)$. $J_0(N)$ seems to increase unboundedly with $N$.}, however, only three stationary points are found: the two global minima $q^\text{s}=(q^\text{s}_1,\dotsc,q^\text{s}_N)$ where all $q^\text{s}_j=\sqrt{6\mu^2/\lambda}$, respectively $-\sqrt{6\mu^2/\lambda}$, and a stationary point of index 1 at $q^\text{s}_j=0$ for all $j$. Moreover, we again find that all stationary values obey $v^\text{s}\leq0$. 

{\em Analytical upper bound on the stationary values.---}The stationary points of the potential $V$ are the solutions of
\begin{equation}\label{eq:statpoint}
\frac{\partial V(q)}{\partial q_k} = \frac{\lambda}{3!}q_k^3 + (4J-\mu^2)q_k - J\sum_{j\in{\mathcal N}(k)}q_j =0.
\end{equation}
Although it is not feasible to solve this set of $N$ coupled nonlinear equations explicitly, the potential energy at a stationary point can be determined by rewriting the potential \eqref{eq:V} in the form
\begin{equation}\label{eq:V2}
V(q)=\sum_{i\in\Lambda}q_i\Biggl[\frac{\lambda}{4!}q_i^3 + \left(\frac{4J-\mu^2}{2}\right)q_i - \frac{J}{2}\sum_{j\in{\mathcal N}(i)}q_j\Biggr].
\end{equation}
Then, substituting \eqref{eq:statpoint} into \eqref{eq:V2}, we obtain the potential energy at a stationary point $q^\text{s}=(q_1,\dotsc,q_N)$,
\begin{equation}\label{eq:Vqs}
V(q^\text{s})= -\frac{\lambda}{4!}\sum_{i\in\Lambda}q_i^4.
\end{equation}
Since $\lambda\geq0$, the potential energy per lattice site at any stationary point is bounded above by zero,
\begin{equation}\label{eq:bound}
v^\text{s}=V(q^\text{s})/N\leq0.
\end{equation}

{\em Comparison with earlier results.---}Our findings, and, in particular, the fact that the stationary values $v^\text{s}$ are nonpositive, disprove earlier results on the relation between thermodynamic phase transitions and stationary points of $V$. These earlier results were phrased in terms of topology changes of certain submanifolds in configuration space, but with the help of Morse theory we can rephrase all statements in terms of stationary points.

In 2004, Franzosi and Pettini announced, and allegedly proved under some conditions on the potential $V$, a necessary condition for a thermodynamic phase transition to occur \cite{FraPe04,*FraPeSpi07}. In the language of stationary points, their claim can be phrased as follows:  If there exists an interval $[a,b]$ such that, for all system sizes $N$ larger than some constant $N_0$, the stationary values $v^\text{s}=V(q^\text{s})/N$ corresponding to the stationary points $q^\text{s}$ of $V$ all lie outside that interval, then in the thermodynamic limit neither a first- nor second-order thermodynamic phase transition can occur at critical potential energies $v_\text{c}\in(a,b)$.

In short, stationary points in the vicinity of some $v_\text{c}$ are claimed to be necessary for a phase transition at $v_\text{c}$. The nearest-neighbor $\phi^4$ potential \eqref{eq:V} satisfies all requirements on $V$ demanded by this theorem. Accordingly, based on the fact that all stationary values are nonpositive \eqref{eq:bound}, the theorem asserts that the critical potential energy of the second-order phase transition of the model cannot be positive. This prediction is in contradiction to the thermodynamic properties of the model and the theorem in \cite{FraPe04,*FraPeSpi07} is falsified by means of a counterexample.

A numerical study of the configuration space topology of the two-dimensional nearest-neighbor $\phi^4$ model was published by Franzosi {\em et al.} in \cite{FraPeSpi00}. The authors reported results for the Euler characteristic $\chi$ (a topological invariant) of the constant-potential energy shells in configuration space, finding a pronounced kink of $\chi$ as a function of $v$ in the vicinity of the transition energy (Fig.\ 3 of \cite{FraPeSpi00}). From the absence of stationary points at positive energies, Morse theory allows one to conclude that $\chi(v)$ is rigorously constant for $v>0$. The kink observed in Fig.\ 3 of \cite{FraPeSpi00} is therefore an artifact of the numerical method used.

\begin{figure}\center
\includegraphics[width=0.48\linewidth]{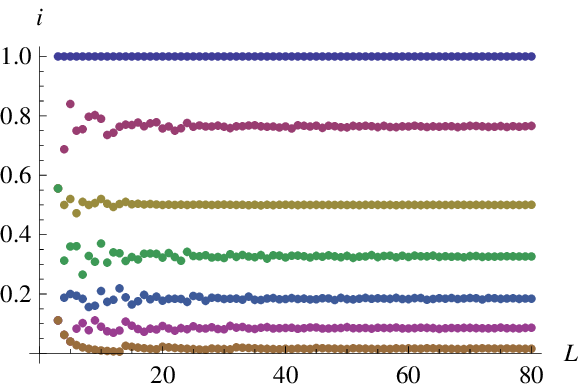}
\hfill
\includegraphics[width=0.48\linewidth]{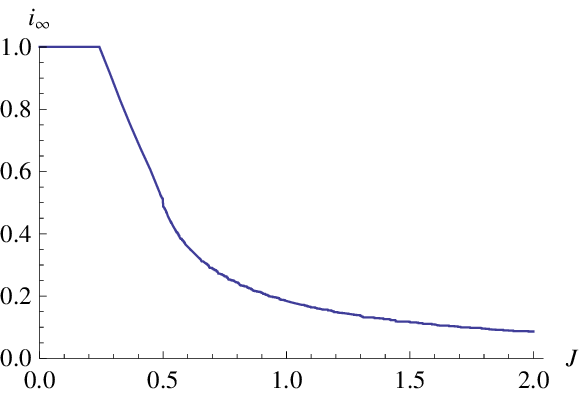}
\caption{\label{fig:00000}
(Color online) Left panel: The relative index $i$ of the stationary point $q^\text{s}=(0,\dotsc,0)$, plotted as a function of the linear system size $L=\sqrt{N}$ for various couplings $J$. From top to bottom: $J=0.1$, 0.36, 0.5, 0.64, 1, 2, 10. Right panel: The large-system limit $i_\infty$ of the relative index, plotted as a function of the coupling constant $J$.
}
\end{figure}

{\em More on the shape of the energy landscape.---}Contrary to the claims in \cite{FraPeSpi00,FraPe04,*FraPeSpi07}, we have seen that the energy landscape in the vicinity of the critical potential energy $v_\text{c}$ of a phase transition can be locally trivial, i.e., free of stationary points with potential energies $v^\text{s}$ in the vicinity of $v_\text{c}$. For $v<0$ and large system sizes, however, when stationary points abound, it is more difficult to explore the shape of the energy landscape. As a first step towards this aim, we study the properties of the stationary point $q^\text{s}=(0,\dotsc,0)$, corresponding to the largest stationary value $v^\text{s}=0$. Its index $I$, i.e., the number of negative eigenvalues of the Hessian matrix of $V$ evaluated at $q^\text{s}$, characterizes the change of the constant-energy shell around $v=0$. With increasing system size $N$, we observe that the relative index $i=I/N$ converges to a finite value, and this value $i_\infty=\lim_{N\to\infty}i/N$ depends on the coupling $J$ (see Fig.\ \ref{fig:00000}). The existence of such a limiting value is a good starting point for discussing the properties of the potential energy landscape in the infinite-system limit. In fact, we can deduce that the potential energy landscape does not approach a simple $N$-dimensional generalization of a double-well potential in the large-system limit, as one might naively have expected. Such a double well, having two degenerate minima at the ground state energy and a stationary point of index $I=1$ at $v=0$, would yield $i_\infty=0$ for the stationary point $q^\text{s}=(0,\dotsc,0)$.

{\em Conclusions.---}By analytical and numerical methods, we have probed certain features of the potential energy landscape of the two-dimensional $\phi^4$ lattice model. The model's phase transition was found to occur at energies well separated from the stationary values $v^\text{s}=V(q^\text{s})/N$ of the potential $V$ (or, equivalently, from topology changes in configuration space). These findings falsify a theorem put forward in \cite{FraPe04,*FraPeSpi07} which claims that stationary points $q^\text{s}$ with stationary values $v^\text{s}=v_\text{c}$ are necessary for a phase transition to occur at a critical potential energy $v_\text{c}$. Since our results imply that the constant-potential energy shells are simply-connected for $v\geq0$, the symmetry breaking phase transition is found to be driven by a concentration-of-measure effect, but is not related to the connectivity of the underlying finite-system energy shells. In other words, even if the integral on the right-hand side of \eqref{eq:se} depends smoothly on $v$, the limiting procedure $N\to\infty$ can, contrary to the claim in \cite{FraPe04,*FraPeSpi07}, destroy this smoothness.

Accordingly, we conclude that the finite-system stationary points provide one possible mechanism of how a phase transition can arise, but not the only one. If the stationary points are at the basis of the transition, the earlier mentioned criterion \cite{KaSchne08,*KaSchneSchrei08} based on the Hessian determinant at stationary points remains valid and can be applied to analyze the phase transition. However, other scenarios are possible and open up interesting perspectives. One possibility is that, even if $V$ has no stationary points when considered on configuration space, it might have such points when treated as a function of $N$ {\em complex} variables. If some of these stationary points, in the thermodynamic limit, approach the real axis and satisfy the Hessian determinant criterion at the same time, they should be capable of inducing a phase transition, despite the absence of stationary points in the real (noncomplex) configuration space. This and other alternative scenarios could open up possibilities for deriving new criteria on the existence or absence of phase transitions, as well as for analytic methods for computing transition energies.

D.\,M.\ acknowledges support by the U.S.\ Department of Energy under Contract No.\ DE-FG02-85ER40237 and by the Science Foundation Ireland Grant No.\ 08/RFP/ PHY1462. M.\,K.\ acknowledges support by the Incentive Funding for Rated Researchers programme of the National Research Foundation of South Africa.

\bibliography{Phi4.bib}

\end{document}